\documentclass[%
preprint,
 amsmath,amssymb,
aps,
prx,
]{revtex4-1}

\usepackage{graphicx,epsfig,epstopdf}
\usepackage{amsmath}
\usepackage{color}
\usepackage{caption}
\usepackage{subcaption}

\def\e{\begin{equation}}
\def\f{\end{equation}}
\def\_#1{{\bf #1}}
\def\.{\cdot}

\def\Re{{\rm Re\mit}}
\def\Im{{\rm Im\mit}}
\def\l#1{\label{eq:#1}}
\def\r#1{(\ref{eq:#1})}

\begin{document}

\title{Perfect Control of Reflection and Refraction Using Spatially Dispersive Metasurfaces
}

\author{V.~S.~Asadchy$^{1,2}$,   
    M.~Albooyeh$^{1}$,
   S.~N.~Tcvetkova$^{1}$,\\ A.~D\'{i}az-Rubio$^{1}$, 
   Y.~Ra'di$^{1,3}$,
     and
   S.~A.~Tretyakov$^{1}$}
 
\affiliation{$^1$Department of Radio Science and Engineering, Aalto University, P.~O.~Box~13000, FI-00076 Aalto, Finland\\
$^2$Department of General Physics, Francisk Skorina Gomel State University, 246019 Gomel, Belarus\\
$^3$Department of Electrical Engineering and Computer Science, University of Michigan, Ann Arbor, Michigan 48109-2122, USA}

\begin{abstract}

Non-uniform metasurfaces (electrically thin composite layers) can be used for shaping refracted and reflected electromagnetic waves. However, known design approaches based on the generalized refraction and reflection laws do not allow realization of perfectly performing devices: there are always some parasitic reflections into undesired directions. In this paper we introduce and discuss a general approach to the synthesis of metasurfaces for full control of  transmitted and reflected plane waves and show that perfect performance can be realized.
The method is based on the use of an equivalent impedance matrix model which connects the tangential field components at the two sides on the metasurface. With this approach we are able to understand what physical properties of the metasurface are needed in order to perfectly realize the desired response. 
Furthermore, we determine the required polarizabilities  of the metasurface unit cells and discuss  suitable cell structures. It appears that only spatially dispersive metasurfaces allow realization of perfect refraction and reflection of incident plane waves into arbitrary directions. In particular, ideal refraction is possible only if the metasurface is bianisotropic (weak spatial dispersion), and ideal reflection without polarization transformation requires spatial dispersion with a specific,  strongly non-local response to the fields. 
  
\end{abstract}

\maketitle

\section{Introduction}

A metasurface is a composite material layer, designed
and optimized in order to control and transform
electromagnetic fields. The layer thickness is negligible
as compared to the wavelength in the surrounding space. 
Conventional devices for wave transformations are either bulky and heavy (e.g., reflector antennas or lenses) or complicated and require active elements  (transmitarray antennas, also called array lenses \cite{Zoya,transmit}). Therefore, it is quite tempting to become able to realize desired transformations (for example, focus or refract wave beams) using extremely thin passive layers.
Recently, there has been considerable interest and progress in creating metasurfaces for controlling reflected and transmitted waves, see recent reviews in \cite{new_review,phil, Caloz_review,Elefth_review, Alu_review}. Some limited manipulations of waves transmitted through a thin metasurface can be accomplished due to specifically designed phase gradient over the metasurface plane \cite{capasso,shalaev,grady,shalaev2}. The required phase gradient is achieved by precise adjustment of the phases of transmitted waves from each metasurface inclusion. Although this approach has enabled  realizations of transmitarrays even at optical frequencies, it suffers from very low efficiency (less than 25\% of transmitted power) and cannot provide control of polarization of the transmitted waves (in fact, it suffers from uncontrollable polarization rotation by $90^\circ$). Subsequently, another approach based on generalized boundary conditions and the use of symmetric metasurfaces was proposed by several researchers \cite{Tailoring,elefth,alu,caloz}. It provided more efficient operation (more than 80\% of transmitted power) and manipulation of polarization \cite{polarization}. However, even this approach cannot ensure ideal performance \cite{passive,Elefth_review}, because these symmetric layers cannot be matched to impedances of two propagating waves (incident and transmitted) in different directions, and, therefore, they inevitably produce some reflections. Most recently, in Refs.~\cite{elefth2,ours} it has been shown that the use of  metasurfaces with asymmetric response can open a possibility to realize metasurfaces for perfect refraction.

Known structures for manipulating reflection (both reflectarrays and metasurfaces) are able to control  reflection phase at each point of the reflector surface and nearly fully reflect the incident power. Representative examples can be found in papers \cite{sun,bozh1,bozh2,mosall2,nanotechnology,prx4,veysi,li}. It has been believed that these properties can allow full control over reflected waves. However, it is not the case. As is shown in the submitted paper \cite{Alu} and in this paper, lossless fully reflecting metasurfaces designed to reflect a plane wave into another plane wave, always produce parasitic beams in undesired directions.
Without proper understanding of the physical properties of metasurfaces responsible for refraction and reflection phenomena it is not possible to create 100\% efficient metasurfaces with desired properties.

Here we address this problem by introducing  a general approach to the design of metasurfaces for arbitrary manipulations of plane waves, both in transmission and reflection. We explain the main ideas of the proposed analytical approach to the synthesis of  general functional metasurfaces using simple but enough general examples of metasurfaces for refraction or reflection of plane waves into arbitrary directions. In the first example, a metasurface between two isotropic  half-spaces (generally different) is designed so that a plane wave incident from one space (the incidence angle $\theta_{\rm i}$) is fully refracted into a plane wave propagating in the second space (the refraction angle $\theta_{\rm t}$), without polarization change. We derive general conditions on the equivalent circuit parameters of the metasurface to ensure perfect refraction while the reflection coefficient is exactly zero (see Section~\ref{conditions}). Subsequently, we consider three different metasurface scenarios to satisfy these conditions (Sections~\ref{sec:teleportation}, \ref{sec:transmitarray}, and \ref{sc3}). The latter scenario was independently considered in \cite{elefth2}.
In the second example, we show how to design metasurfaces which fully reflect plane waves into an arbitrary direction (the reflection angle $\theta_{\rm r}$). In this example, there are two plane waves coexisting in the space in front of the reflecting metasurface. This issue complicates the study, but the solution allows us to approach the problem of synthesis of metasurfaces for the most general field transformations, where the main challenge is to account for interference between multiple plane waves. Indeed, any arbitrary field distribution can be represented as a series of plane waves that interfere on both sides of the metasurface. 
In Section~\ref{sec:power1}, we examine conditions on the metasurface parameters for the perfect reflection regime. Similar conditions were obtained independently in \cite{Alu}. Next, in Sections~\ref{sc4}, \ref{OptDes}, and \ref{sc5}, we consider different scenarios for metasurface realizations.

We show that  perfect control over both refraction and reflection using lossless metasurfaces requires careful engineering of spatial dispersion in the structure. To realize perfect refraction, we need only weak spatial dispersion in form of the artificial magnetism and  bianisotropic omega coupling \cite{serd}. This effect is described by local relations between the exciting electric and magnetic fields and the induced polarizations in the unit cells. Perfect control over reflection using lossless metasurfaces appears to be possible only using strongly non-local metasurfaces: part of the power received in one area of the surface should be ``channelled'' and re-radiated at a different part of the surface. Lossless local-response metasurfaces (that is,  conventional reflectarrays and earlier studied metamirrors) cannot create a perfect reflected plane wave in any direction except the specular and retro directions.  

The results clarify the necessary physical properties of metasurfaces for ideal wave refraction and reflection and explain the limitations of earlier used design methods and earlier  studied realizations in form of electric and  magnetic sheets, inhomogeneous high impedance surfaces and reflectarrays. Possible routes towards realization of ideal and full control over refraction and reflection  are identified and discussed.

\section{Control of transmission: Perfectly refracting metasurface}

\begin{figure}[h!]
\centering
 \epsfig{file=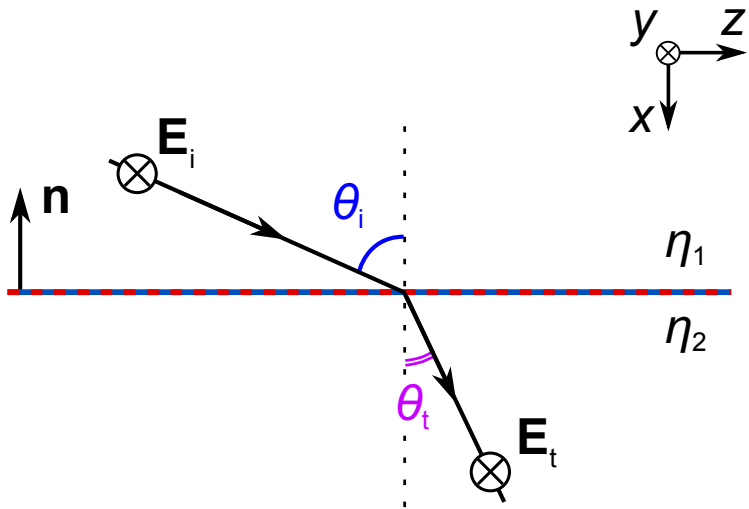, width=0.6\linewidth}  
  \caption{Illustration of the desired performance of an ideally refracting metasurface.}\label{geom}
\end{figure}

As a first step we consider the problem of synthesis of metasurfaces for control of transmitted waves. We require that a given plane wave is fully refracted into another plane wave, without reflections or energy loss. The geometry of the problem is  illustrated in Fig.~\ref{geom}.
The metasurface is located in the $yz$-plane between two isotropic half-spaces with the characteristic impedances $\eta_1$ and $\eta_2$. We assume, without loss of generality of the approach, a transverse electric (TE, with respect to the normal to the surface) incident plane wave. Our approach can be used for waves of arbitrary polarizaitons, including arbitrary polarization transformations, by using the dyadic parameters instead of the scalar ones. 
 
Let us assume that the metasurface is illuminated from medium 1 by a plane wave (with the wavenumber $k_1$ and the electric field vector $\_E_{\rm i}$) at an angle $\theta_{\rm i}$. 
Requiring zero reflections, the tangential field components $\_E_{t1}$ and $\_H_{t1}$  on the illuminated side of the metasurface (at $x=0$) read 
\e \_E_{t1}=\_E_{\rm i}e^{-jk_1\sin\theta_{\rm i}z},\quad \_n\times \_H_{t1}=
\_E_{\rm i}{1\over \eta_1}\cos\theta_{\rm i}  e^{-jk_1\sin\theta_{\rm i}z}, \l{plus} \f
where $z$ is the coordinate along the tangential component of the incident wavevector and the unit vector $\_n$ is orthogonal to the metasurface plane, pointing towards the source. The time-harmonic dependency in form  $e^{j\omega t}$ is assumed.
We want to synthesize a metasurface which will transform this incident wave into a refracted wave
propagating in medium 2 (characterized by parameters $k_2$, $\eta_2$) in some other direction, specified by the angle $\theta_{\rm t}$, without any loss of power. Therefore, the required tangential fields behind the metasurface read
\e \_E_{t2}=\_E_{\rm t}e^{-jk_2\sin\theta_{\rm t}z+j\phi_{\rm t}},\quad \_n\times \_H_{t2}=
\_E_{\rm t}{1\over \eta_2}\cos\theta_{\rm t}  e^{-jk_2\sin\theta_{\rm t}z+j\phi_{\rm t}}. \l{minus}\f For generality, we assume that the refracted wave is phase-shifted by an arbitrary  angle $\phi_{\rm t}$ with respect to the incident plane wave. With these notations, we can choose the origin of the $z$-axis so that both $\_E_{\rm i}$ and $\_E_{\rm t}$ will be real-valued vectors. 

Obviously, the phase of the transmission coefficient 
\e \Phi_{\rm t}(z)=\angle (E_{t2}/E_{t1})=
-k_2\sin\theta_{\rm t}z+\phi_{\rm t}+
k_1\sin\theta_{\rm i}z \l{viki}\f
is not uniform over the surface, as long as $k_2 \sin\theta_{\rm t}\neq k_1 \sin\theta_{\rm i}$. Differentiating the above equation, one can find the 
relation between the incidence and refraction angles in terms of the transmission coefficient phase gradient:
\e k_1\sin\theta_{\rm i}-k_2\sin\theta_{\rm t}={d\Phi_{\rm t}(z)\over dz}.\l{Gen_Snell}\f
This result suggests the simplest approach to the realization of refractive surfaces: Designing a locally-periodical surface whose transmission coefficient is unity in the absolute value (lossless Huygens' sheet) and the phase of the transmission coefficient linearly changes in accordance with  \r{Gen_Snell}. This method was used for a long time in antenna engineering (e.g. \cite{1979}) and more recently in designs of metasurfaces, both in microwaves (e.g. \cite{PRX}) and optics (e.g. \cite{capasso}). However, this approach does not lead to the desired perfect refraction \cite{passive}, and next we will explain how the desired performance can be realized exactly.  

\subsection{Conditions on the equivalent circuit parameters}\label{conditions}

First, let us find the amplitude of the transmitted wave $\_E_{\rm t}$ which corresponds to full power transmission through the metasurface. Looking for possible realizations as metasurfaces with local response,    we equate the normal (to the metasurface) components of the Poynting vector at each point of the metasurfaces, in the two media:
\e {1\over 2}{\rm Re}(\_E_{t1}\times \_H_{t1}^*) = {1\over 2}{\rm Re}(\_E_{t2}\times \_H_{t2}^*), \f
and substitute the field values from \r{plus} and \r{minus}. 
As a result, for metasurfaces with locally full power transmission we obtain
\e \_E_{\rm t}=\_E_{\rm i}\sqrt{\cos\theta_{\rm i}\over \cos\theta_{\rm t}}\sqrt{\eta_2\over \eta_1}.\l{amp_tr}\f
Note that the amplitude of the transmitted wave can be larger or smaller than the amplitude of the incident plane wave, although the metasurface is lossless and the power is conserved in  transmission. This result already tells about a limitation of the mentioned above simple design approach based only on engineering the transmission phase according to \r{Gen_Snell}.

\begin{figure}[h!]
\centering
 \epsfig{file=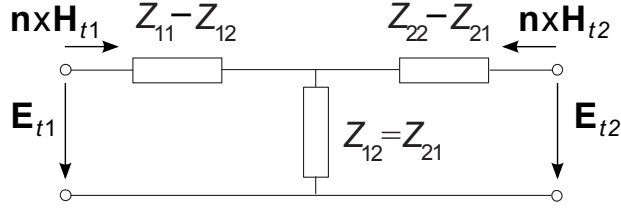, width=0.5\linewidth}  
  \caption{Equivalent $T$-circuit of a  reciprocal metasurface for the considered case of one linear polarization (TE).}\label{T}
\end{figure}

Let us write the linear relations between the tangential fields at the two sides of the metasurface  in form of an impedance matrix: 
\e  \_E_{t1}=Z_{11} \_n\times \_H_{t1} +Z_{12} (-\_n\times \_H_{t2}), \l{11-12} \f
\e  \_E_{t2}=Z_{21} \_n\times \_H_{t1} +Z_{22} (-\_n\times \_H_{t2}),\l{21-22} \f
and find such values of the $Z$-parameters which correspond to this particular field transformation. Knowing the $Z$-parameters of a metasurface, we will be able to determine suitable topologies of constitutive elements (the unit-cell structures) which will realize the desired functionality. Furthermore, the use of the equivalent $T$-circuit (Fig.~\ref{T}) helps in understanding what physical properties the metasurface should have in order to provide the desired response.

Substituting the field values from \r{plus}, \r{minus}, and \r{amp_tr}, we get the following equations for the  $Z$-parameters:
\begin{equation}
\begin{split} e^{-jk_1\sin\theta_{\rm i}z} & = Z_{11}\, {1\over \eta_1}\cos\theta_{\rm i} \, e^{-jk_1\sin\theta_{\rm i}z} \\ 
& -Z_{12}\, {1\over\sqrt{\eta_1 \eta_2}} \sqrt{\cos\theta_{\rm i} \cos\theta_{\rm t}} e^{-jk_2\sin\theta_{\rm t}z+j \phi_{\rm t}},\l{eq1}
\end{split}
\end{equation}
\begin{equation}
\begin{split} e^{-jk_2\sin\theta_{\rm t}z+j \phi_{\rm t}}& =Z_{21}\, {1\over\sqrt{\eta_1 \eta_2}} \sqrt{\cos\theta_{\rm i} \cos\theta_{\rm t}}\, e^{-jk_1\sin\theta_{\rm i}z}\\
& -    Z_{22}\,  {\cos\theta_{\rm t}\over \eta_2}\, e^{-jk_2\sin\theta_{\rm t}z+j \phi_{\rm t}}.\l{eq2a}\end{split}
\end{equation}
Obviously, there are infinitely many solutions for the unknown $Z$-parameters, because we have only two conditions imposed on four complex parameters. Note that solutions with complex values of impedance parameters mean that some of the components  forming the metasurface are either lossy or active, but all these solutions still correspond to the {\it overall lossless} response of the metasurface, because the fields on the two sides of the metasurface form plane waves carrying the same power in both upper and lower half-spaces.

This observation suggests that we can impose some restrictions on the values of the 
equivalent parameters of the metasurface for a specific transformation and achieve different realizations of metasurfaces which all perform the same operation on incident plane waves. The possibility of multiple realizations of arbitrary metasurfaces using the susceptibility model was discussed in Ref.~\cite{caloz}.


\subsection{Teleportation metasurface}
\label{sec:teleportation}

Considering equations \r{eq1} and \r{eq2a}, we observe that while the left-hand sides are single exponential functions (corresponding to either incident or transmitted wave), the right-hand sides are sums of two different exponential functions. This property indicates that in general the $Z$-parameters of the metasurface will depend on the coordinate $z$, that is, the metasurface is, in general, not uniform. However, there is an interesting and conceptually simple solution corresponding to a \emph{homogeneous} metasurface. If we assume that $Z_{12}=Z_{21}=0$, then both equations are satisfied with 
\e Z_{11}={\eta_1\over \cos\theta_{\rm i}},\qquad Z_{22}=-{\eta_2\over \cos\theta_{\rm t}}.\f 
In this scenario, the metasurface is formed by a matched absorbing layer (the input resistance $Z_{11})$, a perfect electric conductor  (PEC) sheet, and an active layer (an ``anti-absorber'' \cite{alutele,tele}) on the other side. The incident plane wave is totally absorbed in the matched absorber. The negative-resistance sheet (resistance $Z_{22}$) together with the wave impedance of medium 2 forms a self-oscillating system whose stable-generation regime corresponds to generation of a plane wave in the desired direction (the refraction angle $\theta_{\rm t}$). Indeed, the sum of the wave impedance of plane waves propagating at the angle $\theta_{\rm t}$ and the input impedance of the active layer is zero, and this is the necessary condition for stable generation.  This structure is similar to the ``teleportation metasurface'' introduced in \cite{alutele,tele} for teleporting waves without changing  the propagation direction. As shown in  \cite{tele}, in  that case if the reflector separating the resistive and active layers is made at least slightly imperfect (parameters $Z_{12}=Z_{21}$ are very small but not exactly zero), the amplitude and phase of the transmitted wave is synchronised with the incident field.

The teleportation metasurface is a theoretically perfect realization of the desired transformation of plane waves in transmission. In particular, when the incidence angle equals  $\theta_{\rm i}$, the reflection coefficient is exactly zero. However, because the input resistance of the metasurface seen from medium 2 is negative, reflections of waves coming from this medium are very strong. Therefore, within the linear model of the negative resistance, the reflection coefficient tends to infinity for waves coming from the direction $\theta_{\rm t}$. Moreover, the amplitude of the field in medium 2 is established due to non-linear saturation of the negative resistance device. Therefore, it is probably practically impossible to ensure that the negative resistance saturates at exactly the desired amplitude of the generated wave. 
Next, we consider an alternative realization, requiring perfect matching of the metasurface for waves coming from medium 2.

\subsection{Transmitarray}
\label{sec:transmitarray}

\begin{figure}[h!]
\centering
 \epsfig{file=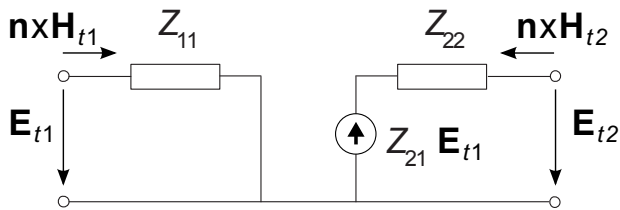, width=0.5\linewidth}  
  \caption{Equivalent $T$-circuit of a nonreciprocal transmitarray realization of refractive  metasurfaces.}\label{T_nonrec}
\end{figure}

Let us  consider alternative realizations demanding that the input impedance of the metasurface seen from medium 2 is matched to the wave impedance in medium 2, so that waves coming from the direction $\theta_{\rm t}$ will not produce any reflections. This requirement can be satisfied if we demand that 
\e Z_{22}={\eta_2\over \cos\theta_{\rm t}}.\f
Now we can find a realization of the metasurface as a nonreciprocal system  where the ideal voltage source in the output branch is defined by 
\e Z_{21}={2\sqrt{\eta_1 \eta_2}\over  \sqrt{\cos\theta_{\rm i} \cos\theta_{\rm t}}}\, e^{-j(k_2\sin\theta_{\rm t}-k_1\sin\theta_{\rm i})z+j \phi_{\rm t}},\f
as follows from \r{eq2a}. 
If the desired response for illumination from medium 1 is the only requirement, we can set $Z_{12}=0$ and $Z_{11}={\eta_1\over \cos\theta_{\rm i}}$, so that for illuminations from medium 1 at the incidence angle $\theta_{\rm i}$ the metasurface is acting as a matched absorber (matched receiving antenna array).
This realization can be modeled by the corresponding nonreciprocal equivalent circuit, shown in Fig.~\ref{T_nonrec}.

This realization reminds conventional transmitarrays \cite{Zoya}. The incident plane wave is received by a matched antenna array on one side of the surface and the wave is launched into medium 2 with a transmitting phase array antenna. In the ideal situation the transmitarray is overall lossless, as   the resistance seen from the illuminated side is in fact the radiation resistance of the transmitting array (the two arrays need to be connected by matched cables). 

The same model describes also the concept of field control and active cloaking using active Huygens' surfaces \cite{elefth,Elefth2}. In that scenario, there is no connection between the receiving side (realized as a matched absorber) and the active array. The incident field is assumed to be known and the amplitudes and phases of sources feeding the radiating array  are set accordingly.

\subsection{Symmetrical double current sheets}
\label{sec:symmetrical}

Within the metasurface paradigm, the simplest approach to realization of refractive metasurfaces is to assume that the refraction is controlled by engineering surface densities of electric and magnetic current sheets, co-existing at the metasurface plane. It is obvious that sheets of \emph{only} electric or \emph{only} magnetic currents cannot offer the desired functionality because of the symmetry of the scattered fields in the forward and backward directions.   
Because electric and magnetic surface current sheets are conveniently modeled by surface impedance relations, it appears reasonable to model refractive metasurfaces by two impedance relations which should hold both for the electric and magnetic surface current densities $\_J_{\rm e}$ and $\_J_{\rm m}$ \cite{Senior,Idemen,Kuester}:
\e  \_J_{\rm e}=\_n\times \_H_{t1}-\_n\times \_H_{t2}=Y_{\rm e}\_E_t=Y_{\rm e}{\_E_{t1}+\_E_{t2}\over 2},\l{Je}\f
\e  \_J_{\rm m}=-\_n\times(\_E_{t1}-\_E_{t2})=Y_{\rm m}\_H_t=Y_{\rm m}{\_H_{t1}+ \_H_{t2}\over 2}.\l{Jm} \f  
Here $\_E_t$ and $\_H_t$ are the averaged tangential electric and magnetic fields at the metasurface plane. Forming sums and differences of  \r{11-12} and \r{21-22}, it is easy to see that relations \r{Je} and \r{Jm} can hold only if the metasurface is symmetric and reciprocal, that is, when $Z_{11}=Z_{22}$  and  $Z_{12}=Z_{21}$. Under these assumptions, 
\e Y_{\rm e}={2\over Z_{11}+Z_{12}}, \qquad Y_{\rm m}=2(Z_{11}-Z_{12}). \l{Yem}
\f 
Since we have only two unknown complex parameters $Z_{11}$ and $Z_{12}$, the solution of \r{eq1} and \r{eq2a} becomes unique and it reads
\e Z_{11}={\eta_1\over \cos\theta_{\rm i}} \frac{e^{-j\Phi_{\rm t}}+e^{j\Phi_{\rm t}}}
{e^{-j\Phi_{\rm t}}-{\eta_1 \cos\theta_{\rm t}\over \eta_2\cos\theta_{\rm i}} e^{j\Phi_{\rm t}}   } ,\l{z_ap2}\f
\e Z_{12}={\sqrt{\eta_1 \eta_2}   \over \sqrt{\cos\theta_{\rm i} \cos\theta_{\rm t}}    }\, {{\eta_1 \cos\theta_{\rm t}\over \eta_2\cos\theta_{\rm i}}+1 \over e^{-j\Phi_{\rm t}}- {\eta_1 \cos\theta_{\rm t}\over \eta_2\cos\theta_{\rm i}}e^{j\Phi_{\rm t}}},\l{z_ap1}\f
where $\Phi_{\rm t}$ is defined by \r{viki}.
We see that these parameters, as well as the electric sheet admittance and magnetic sheet impedance \r{Yem}, are complex numbers, which physically means that the surface is either lossy or active at different values of $z$. For a special case of refraction of a normally incident plane wave at $45^\circ$ such solution for sheet parameters has been published in \cite{Tailoring,circ} and later on discussed in e.g. \cite{Elefth_review}.  

Inspecting \r{z_ap2} and \r{z_ap1}, we see that the metasurface parameters can be purely imaginary for all $z$, corresponding to passive lossless realizations, only if  
\e {\eta_1 \cos\theta_{\rm t}\over \eta_2\cos\theta_{\rm i}}=1,\l{Elefth_condition}\f
in which case
\e 
Z_{12}=j{\eta_1\over \cos\theta_{\rm i}}{1\over \sin \Phi_{\rm t}},\qquad 
Z_{11}=j{\eta_1\over \cos\theta_{\rm i}}\cot\Phi_{\rm t}.\f
Corresponding surface admittances, given by \r{Yem}, are also purely imaginary and coincide with those derived in  \cite{passive} in an alternative way. 
Condition \r{Elefth_condition} physically means that the impedance of the incident plane wave at the top side of the metasurface (${\eta_1\over \cos\theta_{\rm i}}$) equals to the impedance of the refracted wave at the bottom side of the surface (${\eta_2\over \cos\theta_{\rm t}}$). It is, however, in contradiction with the desired field structure: Equations \r{plus} and \r{minus} imply that the ratio of the tangential field components (the wave impedance) must in general change if we require perfect refraction. Thus, lossless double current sheets modeled by impedance relations \r{Je} and \r{Jm} cannot realize perfectly refractive metasurfaces.

In paper \cite{passive} the requirement for equal impedances \r{Elefth_condition} was derived in a different way, demanding the absence of losses, and it was concluded that perfect refraction using lossless metasurfaces was not possible without reflections. Indeed, it is clear that adding some reflected field to \r{plus}, it is possible to make sure that the ratio of the tangential fields on top of the metasurface is the same as at the bottom. This approach is followed nearly in all current literature on lossless metasurfaces for refraction control: Nearly always only symmetric metasurfaces have been considered and used (see \cite{Caloz_review,Elefth_review}) and the realization is thought in form of a symmetric double-current sheet. This is the reason why earlier publications (see the review in \cite{Elefth_review}) state that there must be at least small reflections or there is a need to use active elements. The only known to us exception is the recent paper \cite{elefth2} where the problem is attacked  using the generalized scattering matrix.

Next we show that perfect refraction at an angle which is not equal to the incidence angle is in fact  possible using only lossless structures, but only if the surface is spatially dispersive, exhibiting bianisotropic omega coupling. This result has been independently obtained in \cite{Elefth_new_a}.

\subsection{Metasurface formed by lossless elements}\label{sc3}

In the above example realizations, metasurfaces contained both lossy and active elements, which may require complicated and expensive realizations. It is therefore of interest to consider if and how one can realize the same functionality using only reactive lossless components. 

\subsubsection{Impedance matrix}

To answer this question, we again consider the main set of requirements on the $Z$-parameters of an ideal refractive metasurface 
\r{eq1} and \r{eq2a} and look for a solution where all the $Z$-parameters are purely imaginary (i.e., $Z_{ij}=jX_{ij}$):
\begin{equation}
\begin{split}  e^{-jk_1\sin\theta_{\rm i}z}&= jX_{11}\, {1\over \eta_1}\cos\theta_{\rm i} \, e^{-jk_1\sin\theta_{\rm i}z}\\
& -jX_{12}\, {1\over\sqrt{\eta_1 \eta_2}} \sqrt{\cos\theta_{\rm i} \cos\theta_{\rm t}} e^{-jk_2\sin\theta_{\rm t}z+j \phi_{\rm t}},
\end{split}
\end{equation}
\begin{equation}
\begin{split}  e^{-jk_2\sin\theta_{\rm t}z+j \phi_{\rm t}}& =jX_{21}\, {1\over\sqrt{\eta_1 \eta_2}} \sqrt{\cos\theta_{\rm i} \cos\theta_{\rm t}}\, e^{-jk_1\sin\theta_{\rm i}z}\\
& -    jX_{22}\,  {\cos\theta_{\rm t}\over \eta_2}\, e^{-jk_2\sin\theta_{\rm t}z+j \phi_{\rm t}}.\end{split}
\end{equation}
This is a system of four real-valued equations for four real unknowns $X_{ij}$, which has a unique solution:
\e X_{11}={\eta_1\over \cos\theta_{\rm i}}\cot \Phi_{\rm t} ,\l{x11}\f
\e X_{22}={\eta_2\over \cos\theta_{\rm t}}\cot \Phi_{\rm t}, \l{x22}\f
\e X_{12}=X_{21}={\sqrt{\eta_1 \eta_2}\over  \sqrt{\cos\theta_{\rm i} \cos\theta_{\rm t}} }{1\over \sin\Phi_{\rm t}} .\l{x12} \f
For the case of zero phase shift ($\phi_{\rm t}=0$) formulas \r{x11}--\r{x22} agree with  the result of \cite{elefth2}, obtained using the generalized scattering parameters approach. 

The metasurfaces modeled by \r{x11}--\r{x12} are reciprocal ($X_{12}=X_{21}$). Indeed, the same solution follows from \r{eq1}--\r{eq2a} if we demand that a plane wave coming from the second medium (the incidence angle $\theta_{\rm t}$) is fully transmitted into the first medium in the direction $\theta_{\rm i}$. 
The required physical properties of such metasurfaces 
 can be understood from the corresponding equivalent $T$-circuit (see Fig.~\ref{T}). The circuit is asymmetric, because $X_{11}\neq X_{22}$. This structure of the $Z$-matrix corresponds to bianisotropic omega layers, see a discussion in 
\cite{Joni,AlbOmeg}.
Possible appropriate topologies include arrays of $\Omega$-shaped inclusions~\cite{PRL}, arrays of split rings, double arrays of patches (patches on the opposite sides of the substrate must be different to ensure proper magnetoelectric coupling) \cite{metamirror_AP,absorbers,abs_mohammad,abs_mohammad1}, etc. A more complicated set of three parallel reactive sheets was proposed in \cite{elefth2}. 
 
Previuosly, probably only in paper \cite{elefth2} asymmetric metasurfaces were used for transmission management (equations \r{x11}--\r{x12} also appear in \cite{elefth2} for the case when $\phi_{\rm t}=0$).  Note also that the role of the omega-type bianisotropy of metasurfaces has been discussed in the review paper  \cite{phil}, and omega layers have been  successfully used in single-layer metamirrors \cite{PRL}. 

Comparing to the simple designs based on symmetrical metasurfaces (Section~\ref{sec:symmetrical}), we again see from \r{x11} and \r{x22} that lossless {\it symmetric} realizations with $X_{11}=X_{22}$ are possible only if ${\eta_1\over \cos\theta_{\rm i}}={\eta_2\over \cos\theta_{\rm t}}$, as we already saw from requirement \r{Elefth_condition}. If media $1$ and $2$ are the same, we can conclude that previously proposed symmetrical metasurfaces cannot provide perfect refraction (without parasitic reflections or energy loss).

\subsubsection{Unit-cell polarizabilities and appropriate topologies}\label{UnitCellPol}

Although the impedance matrix model provides a simple tool to design structures for desired wave transformations, it is not directly applicable for identifying  appropriate topologies of the metasurface unit cells. Here we show how to determine what are the required properties of unit cells which realize ideally refractive metasurfaces. Knowing the polarizabilities of each unit cell, we can identify what polarization response should be generated in unit cells and what inclusions are needed to realize this response. 
So-called collective polarizabilities \cite{modeboo} relate the tangential electric and magnetic dipole moments induced in the unit cell to the  fields of the incident wave. 
Knowing the $Z$-parameters of a metasurface is tantamount to knowing  reflection and transmission coefficients. Writing them also  in terms of the collective polarizabilities of unit cells, we can find the required polarizabilities which realize the desired response. 
For the perfect refractive metasurfaces the collective polarizabilities of unit cells read (see \cite{suppl})
\e \widehat{\alpha}_{\rm ee}^{yy}=\frac{S}{j\omega} \frac{\cos{\theta_{\rm i}} \cos{\theta_{\rm t}}}{\eta_1 \cos{\theta_{\rm t}} + \eta_2 \cos{\theta_{\rm i}}}
\left[ 2-\left(\sqrt{\frac{\eta_1 \cos{\theta_{\rm t}}}{\eta_2 \cos{\theta_{\rm i}}}}  + \sqrt{\frac{\eta_2 \cos{\theta_{\rm i}}}{\eta_1 \cos{\theta_{\rm t}}}}  \right) e^{\displaystyle j\Phi_{\rm t}(z)}
\right], \l{aee_tr}\f
\e \widehat{\alpha}_{\rm mm}^{zz}=\frac{S}{j\omega} \frac{\eta_1 \eta_2}{\eta_1 \cos{\theta_{\rm t}} + \eta_2 \cos{\theta_{\rm i}}}
\left[ 2-\left(\sqrt{\frac{\eta_1 \cos{\theta_{\rm t}}}{\eta_2 \cos{\theta_{\rm i}}}}  + \sqrt{\frac{\eta_2 \cos{\theta_{\rm i}}}{\eta_1 \cos{\theta_{\rm t}}}}  \right) e^{\displaystyle j\Phi_{\rm t}(z)}
\right] ,\l{aeemm_tr}\f
\e \widehat{\alpha}_{\rm em}^{yz}=-\widehat{\alpha}_{\rm me}^{zy}=\frac{S}{j\omega} \frac{\eta_2 \cos{\theta_{\rm i}} - \eta_1 \cos{\theta_{\rm t}}}{\eta_1 \cos{\theta_{\rm t}} + \eta_2 \cos{\theta_{\rm i}}},\l{aem_tr} \f
where $S$ is the unit-cell area and $\widehat{\alpha}_{\rm ee}^{yy}$, $\widehat{\alpha}_{\rm mm}^{zz}$, $\widehat{\alpha}_{\rm em}^{yz}$, $\widehat{\alpha}_{\rm me}^{zy}$ are, respectively, electric, magnetic, electromagnetic, and magnetoelectric polarizability components (coupling coefficients). The last two coefficients $\widehat{\alpha}_{\rm em}^{yz}$ and $\widehat{\alpha}_{\rm me}^{zy}$ imply so-called bianisotropic response in the unit cells which models the effect of weak spatial dispersion  \cite{serd}. In other words, the incident electric (magnetic) field should induce also magnetic (electric) polarization in the unit cell. Here, $\widehat{\alpha}_{\rm em}^{yz}=-\widehat{\alpha}_{\rm me}^{zy}$, which is a typical characteristic of reciprocal \textit{omega} inclusions \cite{serd}.

As it can be expected, both the electric and magnetic polarizabilities in \r{aee_tr} and \r{aeemm_tr} depend on $z$, and this dependence is the same for both of them. This result reflects the requirement of zero reflection at any point of the metasurface, which demands the balance of the induced electric and magnetic surface currents at any point (Huygens' condition). On the other hand, the omega coupling coefficient in \r{aem_tr} is constant with respect to $z$ and depends only on the impedances and angles. This result reflects the fact that  bianisotropic coupling of omega-type is necessary to ensure that the waves incident on both sides of the metasurface see the same surface impedance, so that reciprocal full transmission is realized. Since the impedances of the two waves depend only on the impedances of the media and on the two angles, the coupling coefficient also depends only on these parameters. As expected,  we see that when the impedances of the incident and transmitted waves are the same, that is, ${\eta_1\over \cos\theta_{\rm i}}={\eta_2\over \cos\theta_{\rm t}}$, the required coupling coefficient vanishes.

Bianisotropic metasurfaces with the required properties defined by \r{aee_tr}--\r{aem_tr} can be realized as arrays of low-loss particles with the appropriate symmetry. As it was mentioned, for microwave applications, metallic canonical omega particles or double arrays of asymmetric patches can be used. Multilayered topologies were proposed in paper \cite{elefth2}. For optical applications, arrays of properly shaped dielectric particles were introduced as omega-type bianisotropic metasurfaces \cite{Alb1,Alb2}.

It is important to compare the polarizabilities \r{aee_tr}--\r{aem_tr} which are required for realizing  perfect refraction with the polarizabilities found in earlier works on wave transformations in the transmitting regime (e.g.,  \cite{capasso,shalaev,grady,shalaev2,PRX}), where the design approach is based on the geometrical optics model and the ``generalized law of refraction'' \r{viki}. In that theory, the metasurface is assumed to be locally periodical, and the unit cells are designed so that the transmission coefficient has unit amplitude and the desired phase at every point.
These requirements are satisfied if  the electric and magnetic polarizabilities read  (taking the earlier considered special case of normal incidence and identical media at both sides \cite{PRX})  
\e
\widehat{\alpha}_{\rm ee}^{yy}= \frac{1}{\eta^2} \widehat{\alpha}_{\rm mm}^{zz}= {S\over j\omega\eta}\left(1-e^{j\Phi_{\rm t}(z)}\right),\l{old_alphas}\f
and the magnetoelectric coupling coefficient $\widehat{\alpha}_{\rm em}^{yz}$ is zero. 
Periodical arrays formed by unit cells having these collective polarizabilities have unit transmissivity and the transmitted waves have the required phases $\Phi_{\rm t}(z)$, but when the cells are  assembled into a non-uniform array,  the performance becomes non-ideal.  In other words, in order to ensure the desired response of the non-uniform metasurface, properties of periodical arrays formed by its unit cells must deviate from the simple geometrical-optics design recipe  \r{viki}. This result is consistent with that in \cite{Alu_review}.
We can conclude that in order to ensure perfect refraction, it is not enough to make the metasuface bianisotropic (introducing asymmetry with respect to its two sides). 
The electric and magnetic polarizabilities in the exact synthesis [see \r{aee_tr}--\r{aem_tr}] solution are also different  as compared to the conventional synthesis solution \r{old_alphas}.

\section{Control of reflection: Perfectly reflecting metasurface}

In the previous case of refractive metasurfaces, there is only one single plane wave at every point of space. In order to be able to synthesize metasurfaces for general field transformations, we need to understand how to control several plane waves which propagate and interfere in the same space. This problem can be solved at an example of a perfectly reflective metasurface, which we consider next.

\begin{figure}[h!]
\centering
 \epsfig{file=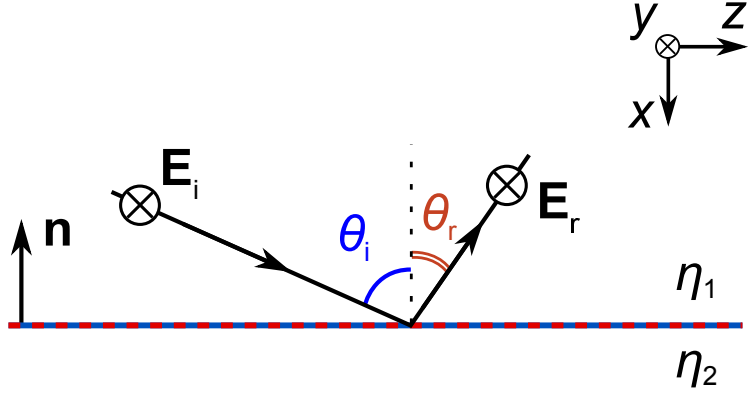, width=0.6\linewidth}  
  \caption{Illustration of the desired performance of an ideally reflecting metasurface. TE incidence is assumed and the metasurface is located in the $yz$-plane.}\label{geom_ref}
\end{figure}

The geometry of the problem is shown in Fig.~\ref{geom_ref}.
The design goal is to fully reflect a plane wave coming from a given direction $\theta_{\rm i}$ into another plane wave propagating in a different and also arbitrary direction $\theta_{\rm r}$. Here, we consider the case when the polarization of the reflected wave is the same as that of the incident wave.  Metasurfaces designed for full reflection were called \emph{metamirrors} in \cite{metamirror_AP,PRL}.
In this scenario, the desired field distribution at the surface of the metamirror is the superposition of two plane waves (the incident wave and the reflected wave):
\begin{equation}
\begin{array}{c}
\vspace*{.3cm}\displaystyle
\_E_{t1}=\_E_{\rm i}e^{-jk_1\sin\theta_{\rm i}z}+\_E_{\rm r}e^{-jk_1\sin\theta_{\rm r}z+j\phi_ {\rm r}},\\ \displaystyle
\quad \_n\times \_H_{t1}=\_E_{\rm i}{1\over \eta_1}\cos\theta_{\rm i}  e^{-jk_1\sin\theta_{\rm i}z}-\_E_{\rm r}{1\over \eta_1}\cos\theta_{\rm r}  e^{-jk_1\sin\theta_{\rm r}z+j\phi_ {\rm r}}.
\end{array}\label{eq:plusmirror}
\end{equation}
Here, $\_E_{t1}$ and $\_H_{t1}$ are the tangential (to the metamirror plane) components of the total electric and magnetic fields at the metamirror surface. 
For generality, we assume that the reflected plane wave can have any desired phase shift $\phi_{\rm r}$ with respect to the incident wave. With these notations, we can choose the origin of the $z$-axis so that both $\_E_{\rm i}$ and $\_E_{\rm r}$ will be real-valued vectors. 
 
Similarly to the refractive metasurface, we see that the phase of the reflection coefficient 
\e \Phi_{\rm r}(z)=-k_1\sin\theta_{\rm r}z+\phi_{\rm r}+
k_1\sin\theta_{\rm i}z\l{mohi}\f
depends on $z$, except the trivial case of specular reflection ($\theta_{\rm i}=\theta_{\rm r}$). Differentiating, we find the relation between the reflection and incidence angles in terms of the gradient of the reflection coefficient phase:
\e k_1(\sin\theta_{\rm i}-\sin\theta_{\rm r})={d\Phi_{\rm r}(z)\over dz}.\l{Gen_refl}\f
Analogously with the transmitting regime, this result suggests a simple design approach: to realize a fully reflective surface (the  amplitude of the reflection coefficient equals unity at each point) but with a linearly varying reflection phase, according to \r{Gen_refl}. Reflecting surfaces with engineered reflection phase are often called high impedance surfaces \cite{HIS} or reflectarrays \cite{encinar}. Such an approach has been used, for example, in \cite{sun,bozh1,bozh2,mosall2,nanotechnology,prx4,veysi,li,PRL}  as well as in all known designs of reflectarrays. 

However, similarly to refracting metasurfaces, in designing reflecting surfaces this simplistic method also does not allow us to exactly realize the desired performance. 
Next,  we present the theory of perfect reflecting surfaces and explore various reflection scenarios, with their advantages and limitations.

\subsection{Power flow into the metamirror}\label{sec:power1}

Applying the same method as in analysing metasurfaces for transmission control, we start from considering the power flow into the metamirror structure. The normal component of the Poynting vector at the reflector surface reads
\e P_n= {1\over 2}{\rm Re}(\_E_{t1}\times \_H_{t1}^*) . \l{smet} \f
Substituting the required field distributions (\ref{eq:plusmirror}), we can write the normal component of the Poynting vector as
\begin{equation}
  P_n={1\over 2\eta_1} \left[ - E_{\rm i}^2 \, \cos\theta_{\rm i} + E_{\rm i} E_{\rm r} \, ( \cos\theta_{\rm r} - \cos\theta_{\rm i}  )\cos\Phi_{\rm r}(z)    +E_{\rm r}^2 \, \cos\theta_{\rm r} \right],
\l{zero_power}
\end{equation}
where the reflection phase $\Phi_{\rm r}(z)$ is defined by \r{mohi}.

If this quantity is identically zero at all points along $z$, the metasurface locally (at every point) acts as a lossless reflector. Conventional realizations of non-uniform reflectors belong to this class of locally responding reflectors.  Examining the above expression, we see that within this scenario, full transformation of an incident plane wave into a single reflected plane wave of the same polarization is impossible, except the cases of specular or retro-reflection, when $\theta_{\rm r}=\pm \theta_{\rm i}$ (this fact is proven also in \cite{Alu}). Indeed, the expression for the normal component of the Poynting vector \r{zero_power} contains an oscillating term, proportional to $\cos\Phi_{\rm r}(z)$, which can be zero only if $\cos\theta_{\rm r} = \cos\theta_{\rm i}$, that is, $\theta_{\rm r}=\pm \theta_{\rm i}$. Therefore, any local, passive and lossless non-uniform reflecting surface will create modulated reflected waves with spatial dependence of the fields different from the design target (\ref{eq:plusmirror}). 

\begin{figure}[h]
  \centering
\epsfig{file=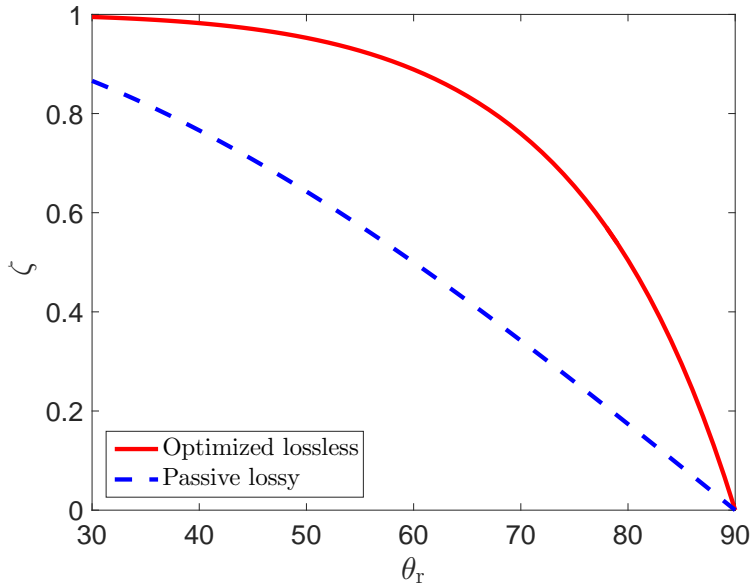, width=0.6\linewidth}
  \caption{Comparison between the power efficiencies of the passive metamirror which reflects a single plane wave [surface impedance \r{Z11_alu_case}, dashed curve] and the optimized metamirror which minimizes reflections into non-desired directions [surface impedance \r{2waves}, solid curve] at normal incidence.}
\label{fig_efficomp}
\end{figure}

The same expression \r{zero_power} tells also that it is possible to reflect a plane wave into only one plane wave along a specified direction if we allow energy loss in the metasurface. To understand this conclusion, let us look for such \emph{constant} amplitude of the reflected wave $E_{\rm r}$ which ensures that $P_n\le 0$ for all $z$ (negative values of $P_n$ corresponds to flow of power into the surface, where it is absorbed). Obviously, this condition can be satisfied if $E_{\rm r}=   E_{\rm i}$, since with this amplitude of the reflected field we have 
\e  
P_n={E_{\rm i}^2\over 2\eta_1} ( \cos\theta_{\rm r} - \cos\theta_{\rm i}  )\left[1+\cos\Phi_{\rm r}(z) \right].\l{alu_case}
\f
Since $1+\cos\Phi_{\rm r}(z)$ is non-negative, $P_n$ is negative or zero at all points of the metasurface $z$ if $\cos\theta_{\rm r} - \cos\theta_{\rm i}\le 0$. This realization scenario was introduced in \cite{Alu}. For instance, if the metamirror is excited by a normally incident plane wave ($\theta_{\rm i}=0$), it is possible to create a single reflected plane wave along any direction, because $\cos\theta_{\rm r}\le 1$ for any $\theta_{\rm r}$. However, as is seen from Eq.~\r{alu_case}, the amount of power which is lost in the metasurface increases with increasing  difference between the incidence and reflection angles. In the limit of $\cos\theta_{\rm r} - \cos\theta_{\rm i}\rightarrow -1$, which corresponds to $\theta_{\rm i}\rightarrow 0$ and $\theta_{\rm r}\rightarrow \pi/2$,  all incident power is completely absorbed. Figure~\ref{fig_efficomp} (dashed line) shows the efficiency of this scenario as a function of the reflection  angle $\theta_{\rm r}$.
The efficiency $\zeta$ is defined as the ratio of the plane-wave power carried into the desired direction $P_{{\rm r}} = {|\_E_{\rm r}|^2\over {2 \eta_1}}  \cos{\theta_{\rm r}}$ to the power of the incident plane wave $P_{{\rm i}} = {|\_E_{\rm i}|^2\over {2 \eta_1}}  \cos{\theta_{\rm i}}$. As it is clear from \r{alu_case} and this figure, increasing the reflection angle results in decreasing the efficiency by a factor of $\cos\theta_{\rm r}/\cos\theta_{\rm i}$ (notice that $E_{\rm r}=   E_{\rm i}$).

Actually, ideal reflection into a single plane wave without losing any power is possible, but only if we allow periodical flow of power into the metamirror structure and back into space. This conclusion is also evident from formula \r{zero_power}. Indeed, we see that if the amplitude of the reflected plane wave equals 
\begin{equation}
E_{\rm r} =\frac{\sqrt{\cos\theta_{\rm i}}}{\sqrt{\cos\theta_{\rm r}}} \, E_{\rm i}, 
\l{nonmodulation}
\end{equation}
the normal component of the Poynting vector is a periodical function with zero average value:
\e 
P_n={E_{\rm i}^2\over 2\eta_1}\,\frac{\sqrt{\cos\theta_{\rm i}}}{\sqrt{\cos\theta_{\rm r}}} \,(\cos\theta_{\rm r}-\cos\theta_{\rm i})\cos \Phi_{\rm r}(z). \l{mod_power}\f
The metasurface performs the desired function perfectly, but the response must be  \emph{strongly non-local}: the power which enters the metasurface structure in the areas where $P_n<0$ must be launched back from the areas where $P_n>0$. Alternatively, the perfect reflection can be achieved if the metasurface has active and lossy elements (being overall lossless in the average over the surface area).
We see again that there is no power flow into the metamirror at any point only if $\theta_{\rm r}=\pm \theta_{\rm i}$, in agreement with the previous conclusion.

\subsection{Required surface impedance}\label{sc4}

Following the introduced synthesis approach based on the impedance matrix, we write 
the linear relation between the tangential fields at the metamirror surface. Assuming that the metamirror is a boundary and the fields behind it are zero ($\_E_{t2}=0$, $\_H_{t2}=0$), we need only one parameter of the $Z$-matrix \r{11-12}--\r{21-22}, the input impedance $Z_{11}$:  
\e  \_E_{t1}=Z_{11}\, \_n\times \_H_{t1} .\l{BCref}\f
Substituting the desired field values from \r{plusmirror}, we get the following equation for the unknown input impedance $Z_{11}$:
\begin{equation}
\begin{split} &\_E_{\rm i}\,e^{-jk_1\sin\theta_{\rm i}z} + \_E_{\rm r}\,e^{-jk_1\sin\theta_{\rm r}z+j\phi_{\rm r}} \\
&=  Z_{11}\, {1\over \eta_1}\left(\_E_{\rm i}\,\cos\theta_{\rm i} \, e^{-jk_1\sin\theta_{\rm i}z}-\_E_{\rm r}\,\cos\theta_{\rm r} \, e^{-jk_1\sin\theta_{\rm r}z+j\phi_{\rm r}}\right).\l{eq1_ref}
\end{split}
\end{equation}

\begin{figure}[h]
  \centering
  \includegraphics[width=.6\linewidth]{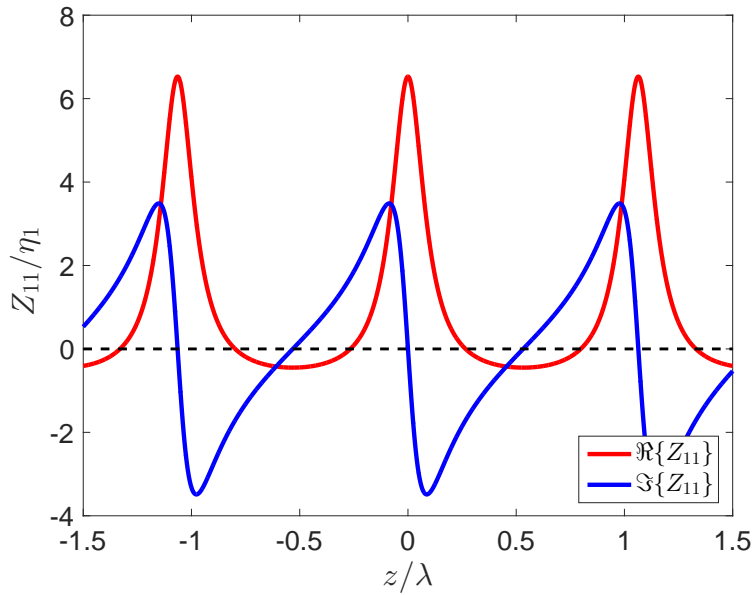}
\caption{The required normalized input impedance $Z_{11}/\eta_1$ of the ideal metamirror for  $\theta_{\rm i}=0^\circ$, $\theta_{\rm r}=70^\circ$, $\phi_{\rm r}=0^\circ$.}
\label{Z11_a}
\end{figure}

For the ideally performing non-local metasurface, which produced the reflected wave with the amplitude given by \r{nonmodulation}, the corresponding input impedance reads
\begin{equation} 
\begin{split} Z_{11}\, = \frac{\eta_1}{\displaystyle \sqrt{\cos\theta_{\rm i}\cos\theta_{\rm r}}} \, \frac{\sqrt{\cos\theta_{\rm r}} + \sqrt{\cos\theta_{\rm i}}\,e^{j\Phi_{\rm r}(z)} }{\displaystyle  \sqrt{\cos\theta_{\rm i}} -\sqrt{\cos\theta_{\rm r}} \, e^{j\Phi_{\rm r}(z)}}.\l{Z_11_const_a}
\end{split}
\end{equation}
We see that the input impedance is a complex number, whose real part is a periodical function of $z$. Figure~\ref{Z11_a} presents the required input impedance  for the case when $\theta_{\rm i}=0^\circ$ and $\theta_{\rm r}=70^\circ$.
The real part of the input impedance periodically takes positive (loss) and negative (gain) values. The surface  \textit{acts}  as if  it is lossy  close to the regions where the reactive impedance is high (close to the regime of a perfect magnetic conductor, PMC) and active in the areas where the reactance is small (close to a perfect electric conductor, PEC). Importantly, this behaviour  does not imply that the surface cannot be passive or lossless. We stress that, on the contrary, properly tuned metasurface with strongly non-local response can emulate such a metamirror: The power which passes through the input surface in the ``lossy'' regions is not absorbed but it is re-radiated from the ``active'' regions. Another possibility to realize the ideal performance dictated by impedance \r{Z_11_const_a} could be a metasurface with truly active and lossy elements where only the overall response is lossless.

\begin{figure}[h]
  \centering
  \includegraphics[width=.6\linewidth]{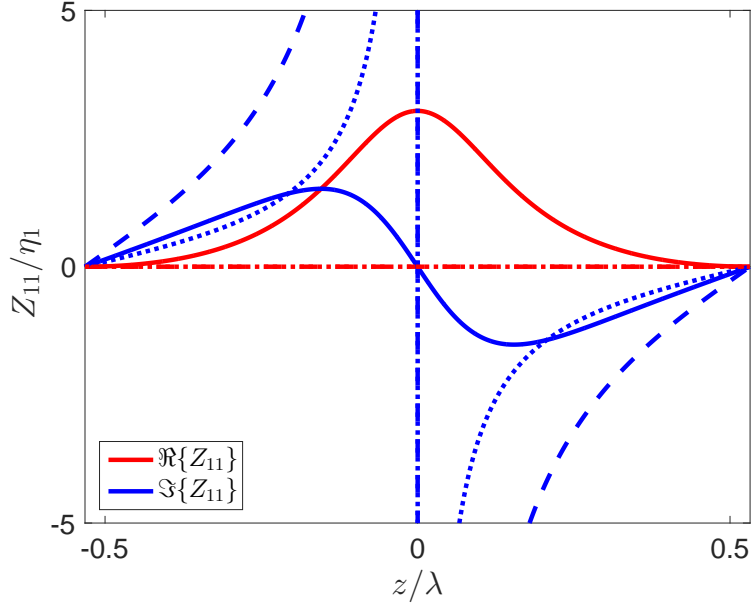}
  \caption{The required normalized input impedance  $Z_{11}/\eta_1$ for passive metamirrors in the case when $\theta_{\rm i}=0^\circ$, $\theta_{\rm r}=70^\circ$, $\phi_{\rm r}=0^\circ$. One period of the metamirror along the $z$-coordinate is shown. The solid, dashed, and dotted lines correspond, respectively, to the lossy metamirrors [Eq.~\r{Z11_alu_case}],  the lossless metamirrors creating two reflected plane waves [Eq.~\r{2waves}], and the conventional non-uniform reflectors [Eq.~\r{viki2}].}
\label{Z11_alu}
\end{figure}

As discussed in Section~\ref{sec:power1}, it is possible to eliminate the need to realize active input impedance (which increases the realization complexities), at the expense of losing some part of the incident power in the metamirror. The surface impedance of such a lossy metasurface, which creates a single plane wave in the desired direction, can be found from \r{eq1_ref} upon substitution of the reflected field amplitude $E_{\rm r}=E_{\rm i}$. The result reads
\e 
Z_{11}=\eta_1{e^{-jk_1\sin\theta_{\rm i}z} + e^{-jk_1\sin\theta_{\rm r}z+j\phi_{\rm r}} \over 
\cos\theta_{\rm i} \, e^{-jk_1\sin\theta_{\rm i}z}-\cos\theta_{\rm r} \, e^{-jk_1\sin\theta_{\rm r}z+j\phi_{\rm r}}} \l{Z11_alu_case}.
\f
An example is plotted as a function of the coordinate in Fig.~\ref{Z11_alu} (the solid lines). As it is seen, the real part of the impedance  is always non-negative, corresponding to the absorbed power given by \r{alu_case}.

So far we have demonstrated that a surface having the input impedance \r{Z11_alu_case} produces a single (non-modulated) reflected wave in the desired direction if the power loss in the metamirror is allowed  [see Fig.~\ref{fig_efficomp}]. However, depending on the application requirements, it can be preferable to allow some modulation of the reflected wave but reduce the power loss. 
In the next section, we present a  scenario in which  the metamirror is lossless at every point, and at the same time  the reflections into non-desired directions are reduced.

\subsection{Optimizing reflections from lossless and local metamirrors}\label{OptDes}
It is possible to optimize the reactance $\Im${$\{Z_{11}\}$} profile of a lossless metamirror in order to minimize reflections into non-desired directions based on particular optimization  criteria. As one example, we notice that there is an interesting lossless design, where all the power which cannot be sent into the desired reflection  direction $\theta_{\rm r}$ is reflected into the specular direction. To demonstrate this possibility, we consider the situation when the difference between the incidence angle $\theta_{\rm i}$ and $\theta_{\rm r}$ is large, so that only three propagating plane waves can exist in the Floquet spectrum of the propagating reflected field  \cite{vardax}
\e
{E}_{\rm r} = \sum_{n=-2}^{0}E_n e^{-jk_1 \left[  (n+1) \sin \theta_{\rm r} - n \sin \theta_{\rm i} \right]z}. \l{Floq}
\f
For the normal illumination ($\theta_{\rm i}=0$), this corresponds to $\theta_{\rm r}$ larger than  $30^\circ$. It is easy to check that a set of three plane waves:  the incident wave,  the wave reflected into the desired direction ($n =0$), and the parasitic plane wave reflected into the specular direction ($n =-1$) exactly satisfy the boundary condition   \r{BCref} with a purely reactive impedance 
\e Z_{11}(z)=j{\eta_1\over \cos\theta_{\rm r}}
\cot\left[\Phi_{\rm r}(z)/2 \right],\l{2waves}
\f
if the wave reflected in the desired direction $\theta_{\rm r}$ is given by 
\e E_{\rm 0}=E_{\rm i}{2\cos\theta_{\rm i}\over \cos\theta_{\rm i}+\cos\theta_{\rm r}},\f
and the wave reflected into the specular direction $\theta_{\rm i}$ is
\e E_{\rm -1}=E_{\rm i}{\cos\theta_{\rm i}-\cos\theta_{\rm r}\over \cos\theta_{\rm i}+\cos\theta_{\rm r}}.\f
The amplitude of the Floquet harmonic $n=-2$ is equal to zero, and the evanescent part of the spectrum also vanishes. These amplitudes have been found by requiring that the normal component of the Poynting vector is identically zero at the surface. In this case, the  metasurface is lossless and exhibits no strong spatial dispersion. Reciprocally, we can conclude that 100\% power reflection in the desired direction can be achieved by illuminating the metasurface by two plane waves at once, properly selecting their relative amplitudes, phases, and propagation directions.

It is interesting that the efficiency of transformation of the incident plane wave into the desired reflected plane wave is much better than for the passive lossy scenario (presented in Section~\ref{sec:power1}) where the parasitic reflections were absent. This conclusion is illustrated in Fig.~\ref{fig_efficomp} by comparing the efficiencies of these two cases.  


The conventional approach for designing lossless non-uniform reflectors is based on the ``generalized reflection law'' \r{Gen_refl}, which corresponds to a linear phase variation along the metasurface. In that approach, the measurface is designed so that the \emph{local} reflection coefficient at every point has unit amplitude and the phase as dictated by \r{Gen_refl}. The local reflection coefficient is defined for an infinite uniform array, that is, the input impedance can be found from  
 \begin{equation}
\_E_{\rm i} + \_E_{\rm r}\,e^{j\Phi_{\rm r}(z)} =  Z_{11}\, {1\over \eta_1}\left(\_E_{\rm i}\,\cos\theta_{\rm i}-\_E_{\rm r}\,\cos\theta_{\rm i} \, e^{j\Phi_{\rm r}(z)}\right),
\end{equation}
where $\_E_{\rm r}=\_E_{\rm i}$ and $\Phi_{\rm r}(z)$ is given by \r{mohi}.
The result reads
\e Z_{11}=j{\eta_1\over  \cos\theta_{\rm i}}\cot\left[\Phi_{\rm r}(z)/2 \right],\l{viki2}
\f
and an example is plotted in Fig.~\ref{Z11_alu}. One can see that the required surface impedance in the conventional reflectors is different from that of the lossless metamirror described by \r{2waves}. In the conventional reflectors, the reflected wave has a complex structure: Generally, several propagating plane waves in different directions and some evanescent fields localized close to the surface are excited. To study the field structure, one can use numerical simulations or the theoretical technique exploited for the case of refraction in \cite{bloch}.

\subsection{Unit-cell polarizabilities and appropriate topologies}\label{PolRef11}

Making use of the boundary conditions on the reflecting metasurface which tell that the tangential electric and magnetic fields are equal, correspondingly, to the surface magnetic and electric current densities, we can find relations between the surface impedance $Z_{11}$ and the collective polarizabilities of unit cells of the metamirror (see \cite{suppl}): 
\e {\eta_1\over \cos{\theta_{\rm i}}}\,{{\cos{\theta_{\rm i}}+\cos{\theta_{\rm r}}}\over {Z_{11}\cos{\theta_{\rm r}}+\eta_1}} = \frac{j\omega}{S} \left({\eta_1\over \cos{\theta_{\rm i}}}\,\widehat{\alpha}_{\rm ee}^{yy}+\widehat{\alpha}_{\rm em}^{yz} \right), \l{pol1}\f
\e Z_{11}\,{{\cos{\theta_{\rm i}}+\cos{\theta_{\rm r}}}\over {Z_{11}\cos{\theta_{\rm r}}+\eta_1}} = \frac{j\omega}{S} \left({\cos{\theta_{\rm i}}\over \eta_1}\,\widehat{\alpha}_{\rm mm}^{zz}-\widehat{\alpha}_{\rm em}^{yz} \right). \l{pol2}\f
Here $S$ is the unit-cell area.
Obviously, these equations have infinitely many solutions for polarizabilities which realize the desired response. The metasurface can be either bianisotropic (omega coupling) or it can be a non-bianisotropic pair of electric and magnetic current sheets. For  the non-bianisotropic realization we set 
\e \widehat{\alpha}_{\rm em}^{yz}=\widehat{\alpha}_{\rm me}^{zy}=0, \f
and find the unique solution
\e \widehat{\alpha}_{\rm ee}^{yy}=\frac{S}{j\omega}\,{{\cos{\theta_{\rm i}}+\cos{\theta_{\rm r}}}\over {Z_{11}\cos{\theta_{\rm r}}+\eta_1}}, \f
\e \widehat{\alpha}_{\rm mm}^{zz}=\frac{S}{j\omega}\, \frac{Z_{11}\,\eta_1}{\cos{\theta_{\rm i}}}\,{{\cos{\theta_{\rm i}}+\cos{\theta_{\rm r}}}\over {Z_{11}\cos{\theta_{\rm r}}+\eta_1}}.\f
We see that in the design of fully reflective metasurfaces, weak spatial dispersion effects are necessary at least in form of the artificial magnetism. If we demand that both magnetic polarizability and the bianisotropy coefficient are zero, the above equations have no solutions. The use of bianisotropy offers additional design flexibilities. 

Knowing the collective polarizabilities required for the desired performance we can immediately see what are the appropriate topologies of unit cells. Since we need both electric and magnetic polarizations, the physical thickness of the reflecting layer must be different from zero, to allow formation of tangential magnetic moments in unit cells. For example, it is not possible to realize the desired performance by any patterning of a single, infinitesimally thin sheet of a perfect conductor. The non-bianisotropic realization scenario suggests the use of a single array of small particles which  are polarizable both electrically and magnetically, such as small metal spirals as in \cite{PRX}. A typical realization based on the bianisotropic route is a high-impedance surface with a PEC ground plane (such as ``mushroom layers'' \cite{HIS}). An important advantage in using bianisotropic effects is the relaxed requirement on the strength of the magnetic response. Especially for optical applications, it is easier to realize strong bianisotropy (which is a first-order dispersion effect) as compared with the artificial magnetism (which is a weaker, second-order effect) \cite{serd}.


\subsection{Ideal metamirrors}\label{sec:power3}

We have seen that all local lossless non-uniform reflectors modulate the reflected waves, which reduces the power efficiency in the desired direction. The operation of conventional planar reflectors (such as high impedance surfaces \cite{HIS}, reflective diffraction gratings \cite{gratings}, and reflectarrays \cite{encinar}) are similar in this respect. Next we discuss the potentials of ideal metamirrors based on non-local and non-reciprocal surfaces.  

As shown above, it is possible to synthesise an \textit{overall} lossless  metamirror which would create an {\it unmodulated} reflected wave into any desired direction, satisfying the requirement 
(\ref{eq:plusmirror}) exactly, with a constant value of the reflected plane wave amplitude $E_{\rm r}$. This goal can be achieved if we require that the normal component of the Poynting vector on the metasurface is zero only in the average over the metamirror period, and not necessarily is equal to zero at every point. In this case, the amplitude of the plane wave reflected into the desired direction is given by \r{nonmodulation}, and the normal component of the Poynting vector oscillates, according to Eq.~\r{mod_power}. 
Realization of such metamirrors requires absorption of power in some areas of the surface and generation of power in some other areas or, alternatively,  power channelling from one area to the other. Conceptually, this scenario of balanced loss and gain can be realized using the same two approaches which were found in the analysis of perfectly refractive metasurfaces: teleportation metasurface (Section~\ref{sec:teleportation}) and transmitarrays (Section~\ref{sec:transmitarray}). 
In the former approach, one can envisage a realization in form of an array of small receiving antennas loaded by positive resistors in the areas where the energy should be partially absorbed, and by negative resistors where the energy should be launched back into space. This arrangement is similar to the teleportation metasurface described in Section~\ref{sec:teleportation}, where such arrays were positioned at the two opposite sides of a metal screen.  
Alternatively, one can envisage a similar array of antennas, where the antennas of the absorbing areas are connected by cables to the antennas of the active areas. Thus, the power received at the absorbing areas is re-radiated by the active areas. It is important to note that both these devices should be non-reciprocal, as the ``active'' antennas should radiate power but not receive it back from space. Actual realization of both these concepts is a challenging task. As to the teleportation approach, one needs non-reciprocal antennas, which can be in principle realized using non-reciprocal materials like magnetized ferrites or using active components. There is also an interesting possibility to use parametric circuits for the same purpose \cite{non_rec_ant}. The non-reciprocal transmitarray approach in principle can be realized also in reflecting metasurfaces, using non-reciprocal circuits inside the metasurface, but it appears that the use of spatial modulation of the surface impedance by external forces (using unit cells equipped with varactors, for example), is more promising.  Conceptually, the desired performance can be achieved by modulating (for example) varactors in all unit cells with the same amplitude but with different phases. Controlling the spatial distribution of the modulation phase, one can possibly realize parameteric amplification or absorption according to the design specifications. Initial work on space-time modulated metasurfaces \cite{space-time,s-t-Caloz} produced interesting and promising results, and we expect that developing this route may lead to realizations of theoretically perfectly operating lossless non-uniform metasurfaces.

\section{Perfectly reflecting polarizers}\label{sc5}
\begin{figure}[h!]
\centering
 \epsfig{file=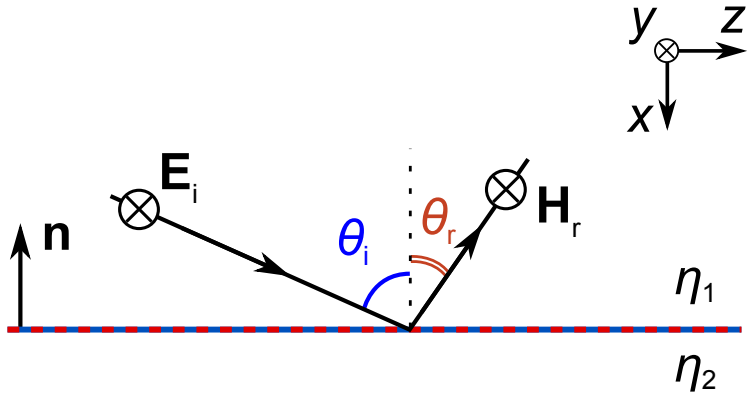, width=0.6\linewidth}  
  \caption{Illustration of the desired performance of an ideal metamirror which perfectly transforms a TE incident wave into a TM reflected wave.}\label{geom3}
\end{figure}

In the previous section, we considered metamirrors which reflect an incident plane wave into a desired direction. However, we encountered either active-lossy realizations of the metamirror or lossless reflection of modulated waves. In this section, we introduce a new solution for a lossless metamirror which ideally reflects the incident wave into the desired direction without any modulations. Since the main reason for modulations of reflected waves is interference between the incident and reflected fields, we construct a metamirror which reflects waves with the polarization orthogonal to that of the incident wave.  As a simple canonical example, we consider the transformation of a transverse electric (TE) wave with the amplitude $E_{\rm i}^{y}$ into a transverse magnetic (TM) wave with the amplitude $E_{\rm r}^{z}=E_{\rm r} \cos{\theta_{\rm r}}$, propagating in the desired direction. Figure~\ref{geom3} shows the problem configuration. 
It is clear that in this case there is no interference between the incident and reflected waves. The desired tangential electric and magnetic fields at the metamirror surface read \e \_E_{t1}=\hat{\_y} \, {E_{\rm i}^{y}}e^{-jk_1\sin\theta_{\rm i}z}+\hat{\_z} \, {E_{\rm r}^{z}}e^{-jk_1\sin\theta_{\rm r}z+\phi_{\rm r}} , \l{plusE} \f and \e \_n\times \_H_{t1}= \hat{\_y} \, {\cos\theta_{\rm i}\over \eta_1} \, {E_{\rm i}^{y}}e^{-jk_1\sin\theta_{\rm i}z}-\hat{\_z} \, {1\over \eta_1 \cos\theta_{\rm r}} \,{E_{\rm r}^{z}}e^{-jk_1\sin\theta_{\rm r}z+\phi_{\rm r}}, \l{plusH} \f
respectively. Considering the metamirror as a boundary, the impedance relation between the tangential electric and magnetic fields \r{11-12}--\r{21-22} in this case reads~\cite{modeboo}
\e  \_E_{t1}=\overline{\overline{Z}}_{11} \cdot \_n\times \_H_{t1}, \l{Z_tens} \f
where the impedance has the matrix form  
\e
\overline{\overline{Z}}_{11} = \begin{bmatrix}
    Z^{yy}_{11}       & Z^{yz}_{11} \\
    & \\
    Z^{zy}_{11}       & Z^{zz}_{11}
\end{bmatrix}.
\f
Notice, in contrast with the previous case, we should consider the full-rank impedance dyadics $\overline{\overline{Z}}$ in order to account for any possible polarization transformation. Substituting \r{plusE} and \r{plusH} into \r{Z_tens}, we obtain the following matrix equation \e
\begin{bmatrix}
    1\\
    \\
    \displaystyle R^{zy}e^{j\Phi_{\rm r}}
\end{bmatrix}=
\begin{bmatrix}
    Z^{yy}_{11}       & Z^{yz}_{11} \\
    & \\
    Z^{zy}_{11}       & Z^{zz}_{11}
\end{bmatrix}
\begin{bmatrix}
    \displaystyle {\cos\theta_{\rm i}\over \eta_1} \\
    \\
    \displaystyle {-1\over \eta_1 \cos\theta_{\rm r}} \,R^{zy}e^{j\Phi_{\rm r}}
\end{bmatrix},
\f
where $R^{zy}={E_{\rm r}^{z}}/{E_{\rm i}^{y}}$ and $\Phi_{\rm r}$ is defined in \r{mohi}. 
The solution of the above equation for the lossless case (i.e., $\Re\{ \overline{\overline{Z}}_{11} \} = 0$) is unique and reads as
\e
\begin{bmatrix}
    Z^{yy}_{11}       & Z^{yz}_{11} \\
    & \\
    Z^{zy}_{11}       & Z^{zz}_{11}
\end{bmatrix}
=j
\begin{bmatrix}
    \displaystyle {\eta_1\over{\cos{\theta_{\rm i}}}}\cot{\Phi_{\rm r}}       & \displaystyle {\eta_1\cos{\theta_{\rm r}}\over{R^{zy}}}{1\over{\sin{\Phi_{\rm r}}}} \\
    & \\
    \displaystyle {\eta_1\over{\cos{\theta_{\rm i}}}}R^{zy}{1\over{\sin{\Phi_{\rm r}}}}       & \displaystyle {\eta_1{\cos{\theta_{\rm r}}}}\cot{\Phi_{\rm r}}
\end{bmatrix}.\l{Z_pol_rot}\f
Here, $Z^{yy}_{11}$ is the metamirror input impedance which is responsible for suppressing unwanted reflections in the specular direction. The proper values of the cross-components $Z^{yz}_{11}$ and $Z^{zy}_{11}$ ensure the polarization rotation, and, finally, $Z^{zz}_{11}$  is responsible for reflection with the orthogonal polarization in the desired direction.

Next, we apply the condition for power conservation (we demand that the normal component of the Poynting vector identically equals zero at the metasurface plane  at each point to ensure local  response) to find the required reflection coefficient  $R^{zy}$. This condition reads
\begin{equation}
   -\left({E_{\rm i}^{y}}\right)^2 \, \cos\theta_{\rm i} \, {1\over 2 \eta_1}+\left({E_{\rm r}^{z}}\right)^2 \, {1\over 2\eta_1\cos\theta_{\rm r}} = 0 ,
\l{zero_powerXpol}
\end{equation}
which defines the reflection coefficient  for the perfect reflection regime:
\e R^{zy}= \sqrt{\cos\theta_{\rm i}\cos\theta_{\rm r}}.\l{Rzy}\f 

As it is clear from \r{Z_pol_rot}, realization of this scenario is  possible with  purely lossless metasurface elements. Moreover, since the reflected field does not interfere with the incident one, there is no field modulation. Therefore, the proposed metamirror provides an ideal and single reflecting wave.

\subsection{Unit-cell polarizabilities and appropriate topologies}\label{PolRef11_rot}

Following the procedure outlined in Sections~\ref{PolRef11}, we can find the relations for collective polarizabilities of unit cells of the proposed metamirror in the case of perfectly reflecting polarizers (see \cite{suppl} 
for details):
\e 1 = \frac{j\omega}{S} \left({\cos{\theta_{\rm i}}\over \eta_1}\,\widehat{\alpha}_{\rm mm}^{zz}+\widehat{\alpha}_{\rm me}^{zy} \right), \l{polrot1}\f
\e R^{zy}e^{j\Phi_{\rm r}} = -\frac{j\omega}{S} \left({\cos{\theta_{\rm i}}\over \eta_1}\,\widehat{\alpha}_{\rm mm}^{yz}+\widehat{\alpha}_{\rm me}^{yy} \right), \l{polrot2}\f
\e {\cos{\theta_{\rm i}}\over \eta_1} = \frac{j\omega}{S} \left({\cos{\theta_{\rm i}}\over \eta_1}\,\widehat{\alpha}_{\rm em}^{yz}+\widehat{\alpha}_{\rm ee}^{yy} \right), \l{polrot3}\f
\e \frac{R^{zy}e^{j\Phi_{\rm r}}}{\eta_1\cos{\theta_{\rm r}}} =- \frac{j\omega}{S} \left({\cos{\theta_{\rm i}}\over \eta_1}\,\widehat{\alpha}_{\rm em}^{zz}+\widehat{\alpha}_{\rm ee}^{zy} \right). \l{polrot4}\f
Obviously, these equations have infinitely many solutions for polarizabilities which realize the desired response.  Even restricting ourselves by reciprocal realizations, the metamirror can be either bianisotropic (both omega and chiral couplings) or it can be  non-bianisotropic with  anisotropic  electric and magnetic responses. Here we show two simple design solutions. In the first design, the metamirror is modeled by anisotropic electric and magnetic polarizabilities. The non-zero polarizabilities read:
\e
\widehat{\alpha}_{\rm mm}^{zz}=  \frac{S}{j\omega} \, \frac{\eta_1}{\cos{\theta_{\rm i}}},
\l{ammzz}\f
\e
\widehat{\alpha}_{\rm mm}^{yz}= - \frac{S}{j\omega} \, \frac{\eta_1}{\cos{\theta_{\rm i}}}\, R^{zy}e^{j\Phi_{\rm r}},
\l{ammyz}\f
\e
\widehat{\alpha}_{\rm ee}^{yy}=  \frac{S}{j\omega} \, \frac{\cos{\theta_{\rm i}}}{\eta_1},
\l{aeeyy}\f
\e
\widehat{\alpha}_{\rm ee}^{zy}= - \frac{S}{j\omega} \, \frac{1}{\eta_1 \cos{\theta_{\rm r}}}\, R^{zy}e^{j\Phi_{\rm r}}.
\l{aeezy}\f
In this design, the bianisotropic properties are excluded, that is, $\widehat{\alpha}_{\rm me}^{zy}= \widehat{\alpha}_{\rm me}^{yy} =\widehat{\alpha}_{\rm em}^{yz} =\widehat{\alpha}_{\rm em}^{zz} =0$. Notice that there is no limitations on the selection of the  polarizability components $\widehat{\alpha}_{\rm ee}^{yz}$, $\widehat{\alpha}_{\rm ee}^{zz}$, $\widehat{\alpha}_{\rm mm}^{zy}$, and $\widehat{\alpha}_{\rm mm}^{yy}$ (they can be chosen from considerations of reciprocity, for example). 

Alternatively, another simple solution of system \r{polrot1}--\r{polrot4} can be found by suppressing the cross polarizability components (i.e., $\widehat{\alpha}_{\rm me}^{zy}=\widehat{\alpha}_{\rm mm}^{yz}=\widehat{\alpha}_{\rm em}^{yz}=\widehat{\alpha}_{\rm ee}^{zy}=0$ ). This implies that the metamirror possesses chiral bianisotropic response:
\e
\widehat{\alpha}_{\rm mm}^{zz}=  \frac{S}{j\omega} \, \frac{\eta_1}{\cos{\theta_{\rm i}}},
\l{ammzz2}\f
\e
\widehat{\alpha}_{\rm me}^{yy}= -\frac{S}{j\omega} \, R^{zy}e^{j\Phi_{\rm r}},
\l{ameyy}\f
\e
\widehat{\alpha}_{\rm ee}^{yy}=  \frac{S}{j\omega} \, \frac{\cos{\theta_{\rm i}}}{\eta_1},
\l{aeeyy2}\f
\e
\widehat{\alpha}_{\rm em}^{zz}= - \frac{S}{j\omega} \, \frac{e^{j\Phi_{\rm r}}}{R^{zy}}\, .
\l{aemzz}\f
while there is no limitation on $\widehat{\alpha}_{\rm ee}^{zz}$, $\widehat{\alpha}_{\rm em}^{yy}$, $\widehat{\alpha}_{\rm me}^{zz}$ and $\widehat{\alpha}_{\rm mm}^{yy}$ (they can be chosen from considerations of reciprocity). It can be shown that if we apply the  reciprocity condition  ($\widehat{\alpha}_{\rm em}^{yy}=-\widehat{\alpha}_{\rm me}^{yy}$ and $\widehat{\alpha}_{\rm me}^{zz}=-\widehat{\alpha}_{\rm em}^{zz}$ \cite{serd}) and  choose
\e
\widehat{\alpha}_{\rm ee}^{zz}=  \frac{S}{j\omega} \, \frac{1}{\eta_1 \cos{\theta_{\rm i}}},
\l{ammzz222}\f
\e
\widehat{\alpha}_{\rm mm}^{yy}=  \frac{S}{j\omega} \, \eta_1 \cos{\theta_{\rm i}},
\l{aemzz222}\f
then the same metamirror dually operates both for TE and TM polarized incident waves. 
One can note a similarity of the conditions on the polarizabilities \r{ammzz2}--\r{aemzz222} with those used earlier for realizing polarization transformers \cite{Niemi} and absorbers \cite{absor11,PRX}. Here we see that  the amplitudes of the polarizabilities should be balanced (as shown in  \cite{Niemi} for the normal incidence), and the ideal reflector operation is ensured by proper adjustments of the chirality parameter phase.

These solutions are only two possibilities, selected for their simplicity.  Other solutions are possible considering \r{polrot1}--\r{polrot4}.

\section{Conclusions and discussions}

In this paper we have introduced a general approach to the synthesis of metasurfaces for arbitrary manipulations of plane waves. We have explained the main ideas of the method on two canonical  examples: A metasurface which perfectly refracts plane waves incident at an arbitrary angle $\theta_{\rm i}$ into plane waves propagating in an arbitrary direction defined by the angle $\theta_{\rm t}$, and a metasurface which fully reflects a given plane wave into an arbitrary direction $\theta_{\rm r}$. The general synthesis approach shows a possibility for alternative physical realizations, and we have discussed different possible device realizations: self-oscillating teleportation metasurfaces, non-local metasurfaces, and metasurfaces formed by only lossless components.  The crucial role of omega-type bianisotropy in the design of lossless-component realizations of perfectly refractive surfaces has been revealed.  

The conventional approach to realization of refractive and reflecting metasurfaces as well as both transmitarray and reflectarray antennas is based on requiring full power transmission or reflection at each point of the surface and providing complete phase control over the transmitted and reflected waves. We have clarified the role of modifications in the required phase gradient for conventional planar refractive/reflective structures in gaining higher efficiencies. Moreover, we have revealed  fundamental limitations of this classical technique and showed how the ideal performance can be realized. For full control over transmission, weak spatial dispersion in form of bianisotropic coupling is necessary, while ideal lossless reflectarray operation calls for the use of structures with a strongly non-local response to the incident fields or structures that transform polarization of reflected waves.

We think that the reason why the role of metasurface bianisotropy in controlling refraction has not been appreciated earlier is that in this field transformation the wave polarization should not change, and it appears natural to expect that bianisotropic effects, such as chirality, are not needed. However, as we have shown here, omega coupling effects, which do not change polarization, are crucial in engineering perfectly matched lossless refractive metasurfaces. 

In contrast to perfectly refracting metasurfaces, creation of perfectly reflecting surfaces  requires careful control over the interference of the incident and reflected waves. We have shown that ideal transformation of an incident plane wave into a reflected plane wave propagating at an angle different from what is dictated by the usual reflection law requires either active structures or passive lossless non-local metasurfaces. We have discussed  the structure of reflected fields and proposed an optimal compromise realization using local and passive metasurfaces.

In the last part of the paper we have shown that the requirement of strong spatial dispersion or  active inclusions   for realization of perfect metamirrors can be lifted if the polarization of the reflected wave is orthogonal to that of the incident field. In this case there is no interference between the incident and reflected wave, and perfect reflection can be realized using only weak spatial dispersion effects (artificial magnetism and chirality), similarly to ideally refractive metasurfaces. 

Since any exciting fields can be expressed in form of a plane-wave expansion, the developed approach can be generalized to metasurfaces for the most general field transformations. We hope that understanding of the physical requirements for perfect metasurface operation in both transmission and reflection regime as well as the developed synthesis method will open a way for design and realization of ultimately thin composite sheets for a broad range of applications, such as lenses, antennas, sensors, etc. 

We would like to note that during the review process of this paper a related preprint \cite{last} has been published, which describes a conceptual realization of perfectly reflecting lossless metasurfaces in form of a set of three parallel reactive sheets. This structure exhibits the required non-local properties (``channeling'' energy in the transverse direction), according to the theory presented here. 

\section*{Acknowledgment}
This work was supported in part by the Academy of Finland (project 287894).

\end{document}